\documentstyle{article}

\begin{document}
\title{Atomic stability and the quantum mass equivalence}
\date{}
\author{Pedro Sancho \\ Centro de L\'aseres Pulsados CLPU \\ Parque Cient\'{\i}fico, 37085 Villamayor, Salamanca,
Spain}
\maketitle
\begin{abstract}
We consider an unexplored aspect of the mass equivalence principle in the
quantum realm, its connection with atomic stability. We show that if
the gravitational mass were different from the inertial one, a
Hydrogen atom placed in a constant gravitational field would become
unstable in the long term. In contrast, independently of the
relation between the two masses, the atom does not become ionized in
an uniformly accelerated frame. This work, in the line of previous
analyses studying the properties of quantum systems in gravitational
fields, contributes to the extension of that programme to internal variables.
\end{abstract}



\section{Introduction}

The formulation of a quantum theory of the gravitational field
remains as one of the main challenges in fundamental physics.
Several authors have signaled the need of a deeper understanding of
the relation between quantum mechanics and gravitation in
order to correctly identify the physical and conceptual roots of the
problem. In particular, we must explore at depth the role of the
mass equivalence, the equality of inertial and gravitational masses,
at the quantum level.

This exploration is part of a more general programme trying to
understand the behavior of quantum systems in gravitational fields.
In a first development, the influential Colella-Overhauser-Werner
(COW) paper showed that terrestrial gravitation acts as an ordinary
force at the level of non-relativistic quantum mechanics
\cite{cow,gre}. Weak gravitational fields can be incorporated into
Schr\"odinger's equation via the usual prescription for the
potential. In addition, the mass equivalence holds in this
experiment. Later, some authors analyzed other aspects of the
problem. For instance, in relation with the universality of free
fall, the quantum time of flight of a particle in a gravitational
field has been computed in \cite{vio,dav}.

In this work we explore an unnoticed aspect of the problem, the
connection existent between atomic stability and the quantum
formulation of the mass equivalence principle. Taking advantage of
the similarities between our problem and the Stark effect we show
that if the inertial and gravitational masses were different a
Hydrogen atom placed in a constant gravitational field would become
ionized in the long term. In other words, the atom in presence of
the field does not have truly stationary solutions
\cite{opp,lan,har}. The solutions obtained by the perturbation
method, which correctly describe the energy-level structure, really
correspond to resonances with a finite lifetime. From a practical
point of view these lifetimes are very long and we do not need to
care about the effect. However, from a fundamental perspective, the
absence of stationary states and their replacement by
quasi-stationary ones is relevant. The atomic stability would be
lost in the long term if the inertial and gravitational masses were
different.

This is not the first time that the behavior of the Hydrogen atom in
gravitational fields has been studied. Modifications of the
energy-level structure have been evaluated in \cite{Par,Pur,Zha} and
even the atom ionization by the action of the field has been
described in \cite{chi}. However, up to our knowledge, these and
other similar analyses have not explored potential connections with
the equivalence principle.

In the second part of the paper we consider the Hydrogen atom from
the point of view of an uniformly accelerated observer in order to
analyze the strong equivalence principle in our problem. We show
that the extended Galilean transformation \cite{gre,gr1},
representing the coordinates change to accelerated frames, does not
lead to ionization processes even in the case of different inertial
and gravitational masses.

Our paper can be viewed as a contribution to the
programme initiated by COW, trying to include, in addition to the
well understood effects of gravity on the wave properties of matter,
some less explored aspects related to the interaction of a
gravitational field with the internal variables. Several papers have
considered some of these aspects. In \cite{gri,es1} the role of
internal variables in studies of the equivalence principle has been
addressed. How the gravitational interaction can lead to decoherence
via the internal degrees of freedom has been studied in \cite{es2}.
Finally, in relation with the compositeness of quantum systems, in
\cite{rus} the author analyzed superpositions of stationary states
of the Hydrogen atom that can violate the equivalence between active
and passive gravitational masses. However, none of these papers has
considered the question of the atomic stability.

The main conclusion of the paper, that the mass equivalence is a
necessary condition for atomic stability in gravitational fields,
shows that that the role of the principle at the quantum level
differs from that at the classical one. We shall consider the
implications of this result in the Discussion.

\section{The atom in a gravitational field}

Let us consider a Hydrogen atom placed in an uniform gravitational
field. We restrict our considerations to weak gravitational fields,
as the terrestrial one, where according to the COW results we can
use the standard form of Schr\"odinger's equation. For our system it
reads
\begin{equation}
i\hbar \frac{\partial \psi}{\partial t}=-\frac{\hbar ^2}{2m_e}
\Delta _{{\bf x}} \psi -\frac{\hbar ^2}{2m_p} \Delta _{{\bf y}}
\psi +V\psi + \bar{m}_e {\bf g} \cdot {\bf x} \psi + \bar{m}_p
{\bf g} \cdot {\bf y} \psi
\end{equation}
where ${\bf x}$ and ${\bf y}$ are respectively the electron and
proton coordinates, $m_e$ and $m_p$ the inertial masses and
$\bar{m}_e$ and $\bar{m}_p$ the gravitational ones. $V=-e^2/|{\bf
x}-{\bf y}|$ is the Coulomb potential and ${\bf g}$ corresponds to the gravitational field intensity.

In order to evaluate the structure of the atom we must separate the
equation into its center of mass and internal parts. The CM and
relative coordinates are given by ${\bf R}=(m_e {\bf x}+m_p {\bf
y})/{M}$ and ${\bf r} = {\bf x}-{\bf y}$, with $M=m_e+m_p$. The
Schr\"odinger equation becomes
\begin{equation}
i\hbar \frac{\partial \psi}{\partial t}=-\frac{\hbar ^2}{2M} \Delta
_{{\bf R}} \psi -\frac{\hbar ^2}{2\mu} \Delta _{{\bf r}} \psi +V\psi
+ \bar{M}{\bf g}\cdot {\bf R} \psi - \frac{\bar{m}_p m_e - \bar{m}_e
m_p}{M} {\bf g}\cdot {\bf r} \psi
\end{equation}
with $\mu =m_em_p/M$ the reduced mass and $\bar{M}=\bar{m}_e +
\bar{m}_p$.

The equation can be separated by introducing the coordinates change
$\psi ({\bf R},{\bf r},t)=\psi _{CM}({\bf R},t) \psi _{rel}({\bf
r},t)$. We obtain the two expressions
\begin{equation}
i\hbar \frac{\partial \psi _{CM}}{\partial t}=-\frac{\hbar ^2}{2M} \Delta
_{{\bf R}} \psi _{CM} +\bar{M}{\bf g}\cdot {\bf R} \psi _{CM}
\end{equation}
and
\begin{equation}
i\hbar \frac{\partial \psi _{rel}}{\partial t}= -\frac{\hbar ^2}{2\mu} \Delta _{{\bf r}} \psi _{rel} +V\psi _{rel} - \frac{\bar{m}_p m_e -
\bar{m}_e m_p}{M} {\bf g}\cdot {\bf r} \psi _{rel}
\end{equation}
When the mass equivalence holds, $m_e =\bar{m}_e$ and $m_p
=\bar{m}_p$, we have $\bar{M}=M$ and $\bar{m}_p m_e = \bar{m}_e m_p$
and we recover the usual equation for both the internal and CM variables. In
contrast, when the masses are different there are non trivial effects.
On the one hand, the CM fall of the atom in the field depends on
$\bar{M}$, which differs from $M$. On the other hand, and more
important, the external field also affects to the internal dynamics.

These internal changes manifest in two different ways: modifying the
energy-level structure and preventing the existence of truly
stationary states. Although the first aspect is not fundamental for
our purposes, we shall briefly discuss it in the next section by the
sake of completeness.

\section{Energy-level structure}

The energy levels are obtained from the time-independent Schr\"odinger equation
\begin{equation}
\hat{H} \psi _{rel}= -\frac{\hbar ^2}{2\mu} \Delta _{{\bf r}} \psi
_{rel} +V\psi _{rel} - {\cal M} {\bf g}\cdot {\bf r} \psi _{rel} = E
\psi _{rel}
\label{eq:ind}
\end{equation}
where, by the matter of simplicity, we have introduced the
notation ${\cal M}M=\bar{m}_p m_e - \bar{m}_e m_p $.

To solve this equation we note its resemblance to that describing
the Stark effect. In both cases we have an atom placed in an
external constant field. As in the Stark effect we resort to
parabolic coordinates, $\xi$, $\eta$ and $\phi$, given by
\cite{gal}: $x_1=\sqrt{\xi \eta} \cos \phi $ , $x_2=\sqrt{\xi \eta}
\sin \phi $ and $x_3=(\xi - \eta )/2$, where $x_3$ is taken as the
direction of the external field. As the external field is very small
when compared to the Coulomb interaction even in strong
gravitational fields, we can invoke a perturbative treatment where
we can decompose the Hamiltonian in the form $\hat{H}=\hat{H}_0 +
\hat{H}_G$ with $\hat{H}_0= -(\hbar ^2/2\mu) \Delta _{{\bf r}} +V $
and $\hat{H}_G=-{\cal M}{\bf g}\cdot {\bf r}$.

The quantum numbers in parabolic coordinates for the unperturbed
problem are denoted as $n_1$, $n_2$ (parabolic quantum numbers) and
$m$ (magnetic quantum number). They are related to the principal
quantum number by the relation $n=n_1+n_2+|m|+1$ \cite{gal}. When
the perturbation is taken into account at first order the energy
becomes
\begin{equation}
E(n,k)=E_0(n) - \frac{3{\cal M}g \hbar }{2 \mu \alpha c}nk
\end{equation}
with $E_0$ the unperturbed energy, $\alpha =e^2/(4\pi \epsilon _0
\hbar c)$ the fine structure constant, $\epsilon _0$ the free space
permittivity, $c$ the light speed and $k=n_1 - n_2$ that takes the
values $k=0, \pm 1, \cdots , \pm (n-1)$. Every level with $n>1$
splits in $2n-1$ equally separated levels. This separation is
proportional to $n{\cal M}g$ \cite{gal}.

The above partial suppression of the degeneracy provides a potential
method to test the mass equivalence in the quantum realm. The
observation of the above splitting in a gravitational field would be
an unequivocal demonstration of the quantum inequality of both
masses. In the absence of an observable splitting, the method
would provide bounds on the possible values of the mass differences.
However, a simple calculation shows that the separation between
sublevels is much smaller than its linewidth.

\section{Atomic stability}

An interesting property of the Stark effect is that in presence of
an external electric field the Hydrogen atom can become unstable.
This property has been presented in the literature in two different
forms. From a physical perspective the electron can tunnel the
potential barrier ionizing the atom \cite{lan}. From a more
mathematical point of view we say that the internal atomic
Hamiltonian has no eigenvalues and the atomic states are no longer
stationary ones \cite{opp,lan,har}. Both approaches are
complementary and lead to the same final result. The ionization of
the atom is the physical consequence of the absence of stationary
states.

A similar property holds for constant gravitational fields. By
similitude with the Stark effect we have that the electron can
tunnel through the potential barrier ionizing the atom. In other
words, the time-independent Schr\"odinger's equation (\ref{eq:ind})
does not have solutions, that is, the operator $\hat{H}_0 +
\hat{H}_G $ has no eigenvalues \cite{opp,har}. This contrasts with
the behaviour of the system in absence of the external field, when
the equation $\hat{H}_0 \psi = E \psi $ has well-known  solutions.
More technically, the spectrum of $\hat{H}$ is continuous while that
of $\hat{H}_0$ is discrete. The solutions to the problem are not
truly bound but correspond to resonances. The perturbative solutions
described in the previous section are good approximations to the
resonances of the system. We can describe the resonance as a
perturbation of a true bound state. The resonances decay after a
very long but finite time to a state of the continuum.

For all practical purposes the time-decay of the resonances in our
problem is very long, specially for the lower energy states. We can
evaluate the corrections to the energy levels, the transition
probabilities, ... in the usual way. However, this property has
important consequences from a more fundamental perspective. We must
resort to a quasi-stationary picture, where bound states are only a
practical approximation. The atom is no longer stable. After a very
long but finite time the atom will lose the electron.

We can estimate the order of magnitude of this time by invoking the
lifetime of the Hydrogen atom in the ground state due to ionization
by the Stark effect \cite{lan}. Replacing the electric field by the
gravitational one we obtain
\begin{equation}
\tau =\frac{{\cal M} g \hbar ^2}{4m_e^3c^5 \alpha ^5} \exp \left(
\frac{m_e^2c^3 \alpha ^3}{{\cal M}g\hbar} \right)
\end{equation}
In statistical terms this time is very long, but with a low
probability we can observe events of this type for much shorter
times.

We conclude that a violation of the mass equivalence would lead to
an unstable behavior of atomic matter in gravitational fields.
Events of this type would only occur with a very low probability,
but they would exist.

\section{Accelerated frames}

In this section we move from gravitational fields to accelerated
observers. It is well-known that the extended Galilean
transformation of coordinates associated with uniform acceleration
\cite{gre,gr1}, introduces a potential into Schr\"odinger's equation
that can be identified with a constant gravitational field. At least
in this formal sense, the strong version of the equivalence
principle (local equivalence of gravitation and acceleration) is
preserved in the quantum realm.

First of all, we derive the Schr\"odinger equation in the
accelerated frame. The coordinates change for this transformation is
named the extended Galilean transformation \cite{gre,gr1}. In the
two-particle case it can be written as
\begin{equation}
{\bf x}={\bf x}' + {\bf Z}(t) \; , \; {\bf y}={\bf y}' + {\bf Z}(t)
\; , \; t=t'
\end{equation}
where ${\bf x}'$ and ${\bf y}'$ are the electron and proton
coordinates in the new frame and ${\bf Z}(t)$ is any function of
time describing a linear displacement. ${\bf Z}(t)$ can also be defined
as the solution of the equation $d^2 {\bf Z}/dt^2= {\bf a}$ with
${\bf a}$ the acceleration of the moving coordinate system. We can
consider any arbitrary translational acceleration, but here we are
only interested into constant ones. Note that every point in the
frame is accelerating at the same rate, so we have a rigid
coordinate system.

As in the one-particle case we express the accelerated wave function
in the form $\psi '({\bf x}', {\bf y}',t') =\exp (i\Phi ({\bf x}',
{\bf y}',t'))\varphi ({\bf x}', {\bf y}',t')$. With the choice
$\hbar \Phi ({\bf x}', {\bf y}',t') =-m_e \dot{\bf Z} \cdot {\bf x}'
-m_p \dot{\bf Z} \cdot {\bf y}' –(M/2) \dot{\bf Z} ^2 t' $ the
Schr\"odinger equation in the accelerated frame reads
\begin{equation}
i\hbar \frac{\partial \varphi}{\partial t'}=-\frac{\hbar ^2}{2m_e}
\Delta _{{\bf x}'} \varphi -\frac{\hbar ^2}{2m_p} \Delta _{{\bf y}'}
\varphi +V\varphi -m_e \ddot{\bf Z} \cdot {\bf x}' \varphi -m_p
\ddot{\bf Z} \cdot {\bf y}' \varphi
\end{equation}
Note that $V=-e^2/|{\bf x}'-{\bf y}'|$, the Coulomb potential, is
invariant under the extended transformation.

Two new terms appear with respect to the inertial equation. When the
equivalence of inertial and gravitational masses holds, we can
identify them with the potential describing the interaction between
the particles and an uniform gravitational field. The direction and
intensity of the field is given by $-\ddot{\bf Z}$, that is, by the
acceleration of the second frame. This result is many times
presented as a quantum (non-relativistic) version of the strong
equivalence principle: at the level of Schr\"odinger's equation an
uniformly accelerated frame is indistinguishable from a constant
gravitational field.

Let us analyze what happens when the mass equivalence does not hold.
The relation $m_e \ddot{\bf Z} \cdot {\bf x}' +m_p \ddot{\bf Z}
\cdot {\bf y}' = M \ddot{\bf Z} \cdot {\bf R}' = (M/\bar{M}) \bar{M}
\ddot{\bf Z} \cdot {\bf R}' $ shows that at variance with a real
gravitational field the acceleration only affects the CM
coordinates. The CM variables behave as those of a composed particle
in a constant gravitational field but with an effective total
gravitational mass $(M/\bar{M}) \bar{M}$ instead of $\bar{M}$. On
the other hand, the internal dynamics is not modified by the
extended Galilean transformation and, consequently, the energy-level
structure does not change with respect to that in an inertial frame.
In particular, there are stationary states and no ionization event
can take place, even if the mass equivalence would be violated.
Atomic matter would be stable in an uniformly accelerated frame even
in these circumstances. We conclude that if inertial and
gravitational masses were not equal the internal dynamics in
accelerated frames and gravitational fields would differ.

\section{Discussion}

We have analyzed the behavior of the Hydrogen atom in constant
gravitational fields and uniformly accelerated frames under the
assumption of different inertial and gravitational masses. The
discussion has focused on the internal properties of the system, a
subject scarcely explored in the context of the equivalence
principle.

Our main conclusion is that the atom can become unstable in these
conditions. The validity of the mass equivalence is a necessary
condition for the stability of atomic systems in gravitational
fields. Classically, the mass equivalence is related to the
universality of free fall. At the quantum level it plays another
role, it guarantees the atomic stability in a gravitational field.

The situation for accelerated observers is completely different. The
violation of the mass equivalence would only affect to the CM
variables. The internal dynamics, in particular the atomic
stability, would remain unaltered with respect to that in an
inertial frame. The mass equivalence does not play any role in the
atomic stability for accelerated observers. This result shows that
there are scenarios where gravitational fields and accelerated
frames are not equivalent, a property not in the spirit of the
equivalence principle. Note, however, that although the mass
equivalence is irrelevant for the stability, its validity is a
necessary condition for the strong equivalence to hold in the
quantum realm (in the specific sense related to the Galilean
transformations) because as discussed in the previous section the
{\it fictitious gravitational potential} appearing in the equations
after the Galilean transformation depends on $M$ instead of
$\bar{M}$.

Our analysis also highlights the fact that a complete understanding
of the mass equivalence in the quantum realm demands the
consideration of the internal degrees of freedom. This agrees with
other studies showing that the internal variables must be taken into
account in order to get a full understanding of the behavior of
quantum systems in gravitational fields \cite{gri,es1,es2}.

The capacity of the gravitational potential (when the mass
equivalence is violated) to modify the atomic spectra and the
stability conditions does not depend on its intensity. Even a tiny
potential is enough to transform a discrete spectra into a
continuous one \cite{lan,har}. This is a qualitative, rather than
quantitative, characteristic of the effect .

We have restricted our considerations to the COW- or weak
gravitational field-regime, where non-relativistic quantum mechanics
describes the behavior of the atom. Even with this limitation our
approach can be relevant to study the interplay between gravitation
and quantum theory. Note, at this respect, that it has been proposed
to use the non-relativistic COW experiment as a low-energy window to
look into the structure of space-time \cite{ani}. The
non-relativistic approach notoriously simplifies the mathematical
treatment of the problem. In order to complete the analysis of the
problem we must consider relativistic quantum mechanics to see if
the same results hold in the new regime. We can consider the
relativistic corrections to the Hydrogen atom or resort directly to
the Dirac equation. In any of the two cases, these considerations
would enlarge too much the paper and must be analyzed separately. As
signaled in the Introduction several authors have studied the
Hydrogen atom in a gravitational field, both in the relativistic
\cite{Par,Pur,Zha} and non-relativistic \cite{chi} regimes (although
none of them in connection with the equivalence principle). The
relativistic considerations indicate that one must analyze in detail
the decomposition into CM and internal variables \cite{Pur,Zha}.

\end{document}